\documentclass[useAMS,usenatbib]{mnras}
\usepackage{multirow}
\usepackage{epsfig}
\usepackage{subfigure}
\usepackage{graphicx}
\usepackage{amsmath,amssymb}

\makeatletter
\let\jnl@style=\rm
\def\ref@jnl#1{{\jnl@style#1}}

\def\aj{\ref@jnl{AJ}}                   
\def\araa{\ref@jnl{ARA\&A}}             
\def\apj{\ref@jnl{ApJ}}                 
\def\apjl{\ref@jnl{ApJ}}                
\def\apjs{\ref@jnl{ApJS}}               
\def\ao{\ref@jnl{Appl.~Opt.}}           
\def\apss{\ref@jnl{Ap\&SS}}             
\def\aap{\ref@jnl{A\&A}}                
\def\aapr{\ref@jnl{A\&A~Rev.}}          
\def\aaps{\ref@jnl{A\&AS}}              
\def\azh{\ref@jnl{AZh}}                 
\def\baas{\ref@jnl{BAAS}}               
\def\jrasc{\ref@jnl{JRASC}}             
\def\memras{\ref@jnl{MmRAS}}            
\def\mnras{\ref@jnl{MNRAS}}             
\def\pra{\ref@jnl{Phys.~Rev.~A}}        
\def\prb{\ref@jnl{Phys.~Rev.~B}}        
\def\prc{\ref@jnl{Phys.~Rev.~C}}        
\def\prd{\ref@jnl{Phys.~Rev.~D}}        
\def\pre{\ref@jnl{Phys.~Rev.~E}}        
\def\prl{\ref@jnl{Phys.~Rev.~Lett.}}    
\def\pasp{\ref@jnl{PASP}}               
\def\pasj{\ref@jnl{PASJ}}               
\def\qjras{\ref@jnl{QJRAS}}             
\def\skytel{\ref@jnl{S\&T}}             
\def\solphys{\ref@jnl{Sol.~Phys.}}      
\def\sovast{\ref@jnl{Soviet~Ast.}}      
\def\ssr{\ref@jnl{Space~Sci.~Rev.}}     
\def\zap{\ref@jnl{ZAp}}                 
\def\nat{\ref@jnl{Nature}}              
\def\iaucirc{\ref@jnl{IAU~Circ.}}       
\def\aplett{\ref@jnl{Astrophys.~Lett.}} 
\def\apspr{\ref@jnl{Astrophys.~Space~Phys.~Res.}}
\def\bain{\ref@jnl{Bull.~Astron.~Inst.~Netherlands}}
\def\fcp{\ref@jnl{Fund.~Cosmic~Phys.}}  
\def\gca{\ref@jnl{Geochim.~Cosmochim.~Acta}}   
\def\grl{\ref@jnl{Geophys.~Res.~Lett.}} 
\def\jcp{\ref@jnl{J.~Chem.~Phys.}}      
\def\jgr{\ref@jnl{J.~Geophys.~Res.}}    
\def\jqsrt{\ref@jnl{J.~Quant.~Spec.~Radiat.~Transf.}}
\def\memsai{\ref@jnl{Mem.~Soc.~Astron.~Italiana}}
\def\nphysa{\ref@jnl{Nucl.~Phys.~A}}   
\def\physrep{\ref@jnl{Phys.~Rep.}}   
\def\physscr{\ref@jnl{Phys.~Scr}}   
\def\planss{\ref@jnl{Planet.~Space~Sci.}}   
\def\procspie{\ref@jnl{Proc.~SPIE}}   

\makeatother

\newcommand{\hst}{{\it HST}}

\title[\hst\ unveils a relativistic Broad Line Region in NGC~3147]{\hst\ unveils a compact mildly relativistic Broad Line Region in the candidate true type 2 NGC~3147}

\author[Stefano Bianchi, et al.]{Stefano Bianchi$^1$\thanks{E-mail: bianchi@fis.uniroma3.it (SB)}, Robert Antonucci$^2$, Alessandro Capetti$^3$, Marco Chiaberge$^{4,5}$,
\newauthor Ari Laor$^6$, Loredana Bassani$^7$, Francisco J. Carrera$^8$, Fabio La Franca$^1$,   
\newauthor Andrea Marinucci$^1$, Giorgio Matt$^1$, Riccardo Middei$^1$, Francesca Panessa$^9$\\
$^1$Dipartimento di Matematica e Fisica, Universit\`a degli Studi Roma Tre, via della Vasca Navale 84, 00146 Roma, Italy\\
$^2$Department of Physics, University of California, Santa Barbara, CA, USA\\
$^3$INAF - Osservatorio Astrofisico di Torino, Via Osservatorio 20, I-10025 Pino Torinese, Italy\\
$^4$Space Telescope Science Institute, 3700 San Martin Dr., Baltimore, MD 21210, USA\\
$^5$Johns Hopkins University, 3400 N. Charles Street, Baltimore, MD 21218, USA\\
$^6$Physics Department, Technion - Israel Institute of Technology, Haifa 32000, Israel\\
$^7$INAF/IASF Bologna, Via P. Gobetti 101, I-40129 Bologna, Italy\\
$^8$Instituto de F\'isica de Cantabria (CSIC-Universidad de Cantabria), E-39005 Santander, Spain\\
$^9$INAF Istituto di Astrofisica e Planetologia Spaziali, via Fosso del Cavaliere 100, I-00133 Roma, Italy\\
}
\begin{document}

\maketitle

\label{firstpage}

\begin{abstract}
NGC~3147 has been considered the best case of a true type 2 AGN: an unobscured AGN, based on the unabsorbed compact X-ray continuum, which lacks a broad line region (BLR). However, the very low luminosity of NGC 3147 implies a compact BLR, which produces very broad lines, hard to detect against the dominant background host galaxy. Narrow ($0.1\arcsec \times0.1\arcsec$) slit \hst\ spectroscopy allowed us to exclude most of the host galaxy light, and revealed an H$\alpha$ line with an extremely broad base (FWZI$\sim27\,000$ km s$^{-1}$). The line profile shows a steep cutoff blue wing and an extended red wing, which match the signature of a mildly relativistic thin accretion disk line profile. It is indeed well fit with a nearly face on thin disk, at $i\sim23\degr$, with an inner radius at $77\pm15$ r$_g$, which matches the prediction of $62^{+18}_{-14}$ r$_g$ from the $R_{\rm BLR} \sim L^{1/2}$ relation. This result questions the very existence of true type 2 AGN. Moreover, the detection of a thin disk, which extends below 100 r$_g$ in an $L/L_{\rm Edd}\sim10^{-4}$ system, contradicts the current view of the accretion flow configuration at extremely low accretion rates.
\end{abstract}


\newcommand{\JS}[1]{{\color{red}{#1}}}
\newcommand{\nH}{n_{\rm H}}
\section{Introduction}

The polarized flux (scattered light) spectra of many type 2 Seyfert galaxies show a Big Blue Bump and broad emission lines, showing the presence of a hidden Seyfert 1-like nucleus.  Because the percent polarization is very high ($\gtrsim15\%$), and approximately \textit{perpendicular} to the radio structure axis\footnote{That excludes interstellar dichroic absorption, and applies to the percent polarization of the scattered light alone \citep{Antonucci2002}.}, it was inferred that photons can only escape from the nuclear regions by travelling \textit{along} the axis and then scattering into the line of sight; escape along the equatorial latitudes is blocked by a `dusty torus' or similar structure. This leads to the Unified Model \citep[UM,][]{Antonucci1993,Antonucci2012}, the simplest version of which states that the spectral types differ solely due to inclination of the axis with respect to the line of sight. The nucleus is seen directly when the inclination is small.

However, a significant fraction of Seyfert 2 galaxies do not show a hidden broad-line region (BLR), even in high-quality spectro-polarimetric data (\citealp{Tran2001,Tran2003}, but see the smaller fraction in \citealp{RamosAlmeida2016}). A possible explanation is that any mirror reflecting the broad lines has low scattering efficiency 
or is obscured \citep{Miller1990,Heisler1997}. The lack of polarized broad lines appears also to be associated with a stronger dilution from the host galaxy or from a circumnuclear starburst, making their detection more challenging \citep{Alexander2001,Gu2001}.  Nevertheless, the X-ray emission often provides clear evidence for a buried type 1 AGN, as indicated by the detection of a highly absorbed, or reflection-dominated, X-ray continuum \citep[e.g.][]{Marinucci2012a}.

Some low luminosity Seyfert 2s show instead an unobscured X-ray continuum, strongly suggesting a direct view of the innermost regions close to the massive black hole (BH) (as much as 30\% of the entire type 2 population, according to \citealp{Panessa2002,Netzer2015}, but see the much lower fraction suggested by \citealp{Malizia2012}). Furthermore, in some of these objects the optical and/or the X-ray continuum varies rapidly, clearly excluding that we are instead observing large-scale reflection of the inner nucleus \citep[e.g.][]{Hawkins2004,Bianchi2012,Bianchi2017a}. Are these \textit{true} type 2 AGN, intrinsically lacking a BLR?  Does the AGN unification break down in low luminosity AGN?

A plausible model for the origin of the BLR gas is a disc wind, driven by radiation pressure on UV resonance lines \citep[][]{Shlosman1985,Murray1995,Proga2004}, on dust grains \citep[][]{Czerny2011,Baskin2018a}, or magnetically driven \citep[][]{Emmering1992,Lovelace1998}. This wind is expected to disappear for low luminosities, or accretion rates \citep{Nicastro2000,Laor2003,Elitzur2006a,Elitzur2009}. Indeed, the true type 2 best candidates have low Eddington ratios: NGC~3147 \citep[$\log L_\mathrm{bol}/L_\mathrm{Edd}\simeq-4$:][]{Bianchi2008c}, Q2131-427 \citep[$\log L_\mathrm{bol}/L_\mathrm{Edd}\simeq-2.6$:][]{Panessa2009b}, NGC~3660 \citep[$\log L_\mathrm{bol}/L_\mathrm{Edd}\simeq-2$: ][]{Bianchi2012}. These are unambiguous candidates since that the lack of the optical broad lines was found together with simultaneous X-ray observations which certify an absorption-free line of sight to the nucleus. 

Have the above theoretical predictions, that the BLR disappears at very low luminosities/accretion rates, been vindicated by these objects? One needs to be careful as the optical spectra of low $L_\mathrm{bol}/L_\mathrm{Edd}$ AGN are heavily dominated by the host galaxy emission. It is extremely difficult to set tight upper limits on the flux of any broad lines and their equivalent widths \textit{relative to any nuclear continuum} in the heavily starlight-dominated ground-based spectra \citep[see][for a comprehensive criticism on true type 2s]{Antonucci2012}. Furthermore, if the BLR does exist, it must be very compact, since the size of the BLR scales as $L_\mathrm{bol}^{1/2}$ for $L_{\rm bol}$ in the range of $10^{40}$-$10^{47}$ erg s$^{-1}$ \citep[e.g.][]{Kaspi2007,Bentz2013}. This relation is naturally explained by assuming that the outer boundary of the BLR is determined by the dust sublimation radius, which is set by the continuum luminosity \citep[][]{Laor1993,Netzer1993a}. The compact size leads to a high Keplerian velocity in the BLR, which can make the lines extremely broad, and thus even harder to detect. Indeed, narrow slit \hst\ observations revealed very broad H$\alpha$ lines in a number of low luminosity AGN \citep{Ho2000b,Shields2000,Barth2001b,Balmaverde2014b}.

NGC~3147 \citep[z=0.009346:][]{Epinat2008} is the best of the three true type 2 Seyfert candidates with simultaneous optical and X-ray observations \citep{Bianchi2012}, being the one with the most extensive X-ray and optical coverage. Its X-ray spectrum shows very tight constraints on the column density of any obscuring gas along its line-of-sight, and a modest fraction of reprocessing components (iron K$\alpha$ line and Compton hump) even at energies larger than 10 keV, thus excluding with high confidence that we are observing the reflected component of a Compton-thick source \citep{Panessa2002,Bianchi2008c,Matt2012,Bianchi2017a}. Significant X-ray variability on time-scales as short as weeks \citep[][]{Bianchi2017a}, adds further support to a direct view of a compact emitting region.
\citet{Bianchi2008c} analysed the (galaxy-subtracted) optical spectrum taken at the Observatorio de Sierra Nevada simultaneously with the \textit{XMM-Newton} observation, reporting an upper limit to a broad (Full Width at Half Maximum, FWHM, fixed to 2000 km s$^{-1}$) component of the H$\alpha$ line corresponding to a luminosity of $2\times10^{38}$ erg s$^{-1}$. This is significantly lower than the value of $\simeq10^{40}$ erg s$^{-1}$, expected from the X-ray luminosity, based on the relation followed by type 1 AGN \citep{Stern2012}. This is also true for the $\mathrm{L_{H\beta}^{broad}/\nu L_\nu^{1.4\,GHz}}$ and $\mathrm{L_{H\alpha}^{broad}/L_\nu^{10\,\mu m}}$ ratios \citep{Shi2010a}. 

Is this enough to classify it as a true type 2 AGN? The above-mentioned limits on any broad component are valid only if the width and profile of the line are not extreme. The only way to definitely exclude the predicted BLR emission is narrow slit \textit{HST} spectroscopy, which allows for a clear detection of the low luminosity AGN emission. Here we present a \textit{HST} observation of NGC~3147, designed to exclude the bulk of the host emission, thanks to its narrow slit ($0.1\arcsec$).
\vspace{-0.5cm}
\section{Observation and Data Reduction}

The target was observed with the {\it Space Telescope Imaging Spectrograph} (STIS) on board the {\it Hubble Space Telescope} on May 17th 2018. The observations were performed using the CCD with the $52\arcsec \times0.1\arcsec$ slit and
the G750L low resolution grating. The pointing aimed at the host galaxy photocenter, and automatically re-centered to the location of the brightest visible source in the acquisition image. 
The image shows that the AGN is clearly visible as a point source at the center of the field of view. Four more spectra in contiguous regions were also taken, in case the automatic centering procedure did not accurately fit the AGN in the $0.1\arcsec$ slit. Each spectrum was taken with a total exposure time of 410s, split into two CR-SPLIT exposures of equal duration to remove cosmic ray events. The automatic re-centering procedure was able to accurately place the target at the center of the slit.

We retrieved the raw data and the calibration files from the Mikulski Archive for Space Telescopes (MAST). We processed the data to correct for charge transfer inefficiency (CTI) using the Python script provided by the STScI STIS team, based on the model by \citet{Anderson2010}. In addition to providing a pixel-based CTI correction for both the dark frame and each of the raw images, the script also performs the basic calibration steps (bias and dark subtraction and flat field correction) and combines the two CR-SPLIT images into a single 2D spectral image.
The nuclear spectrum was rectified and extracted from a 2 pixel aperture ($0.1\arcsec$) using the {\it PyIRAF} task {\it splot}. We used a synthetic PSF obtained with TinyTim to derive the appropriate aperture correction from 2 pixel to the standard 7 pixel aperture.

The spectrum was analysed with \textsc{Xspec} 12.10.1 \citep[][]{Arnaud1996}, and the fitting model convolved with a Gaussian smoothing with $\sigma=3.14$\AA, to match the instrumental resolution\footnote{http://www.stsci.edu/hst/stis/performance/spectral\_resolution}. In the following, (statistical only) errors and upper limits correspond to the 90 per cent confidence level for one interesting parameter, apart from the plots, where $1\sigma$ error bars are shown. The adopted cosmological parameters are $H_0 = 70$ km s$^{-1}$ Mpc$^{-1}$, $\Omega_\Lambda = 0.73$ and $\Omega_m = 0.27$.

\section{Spectral Analysis}

\begin{figure}
\includegraphics[width=\columnwidth]{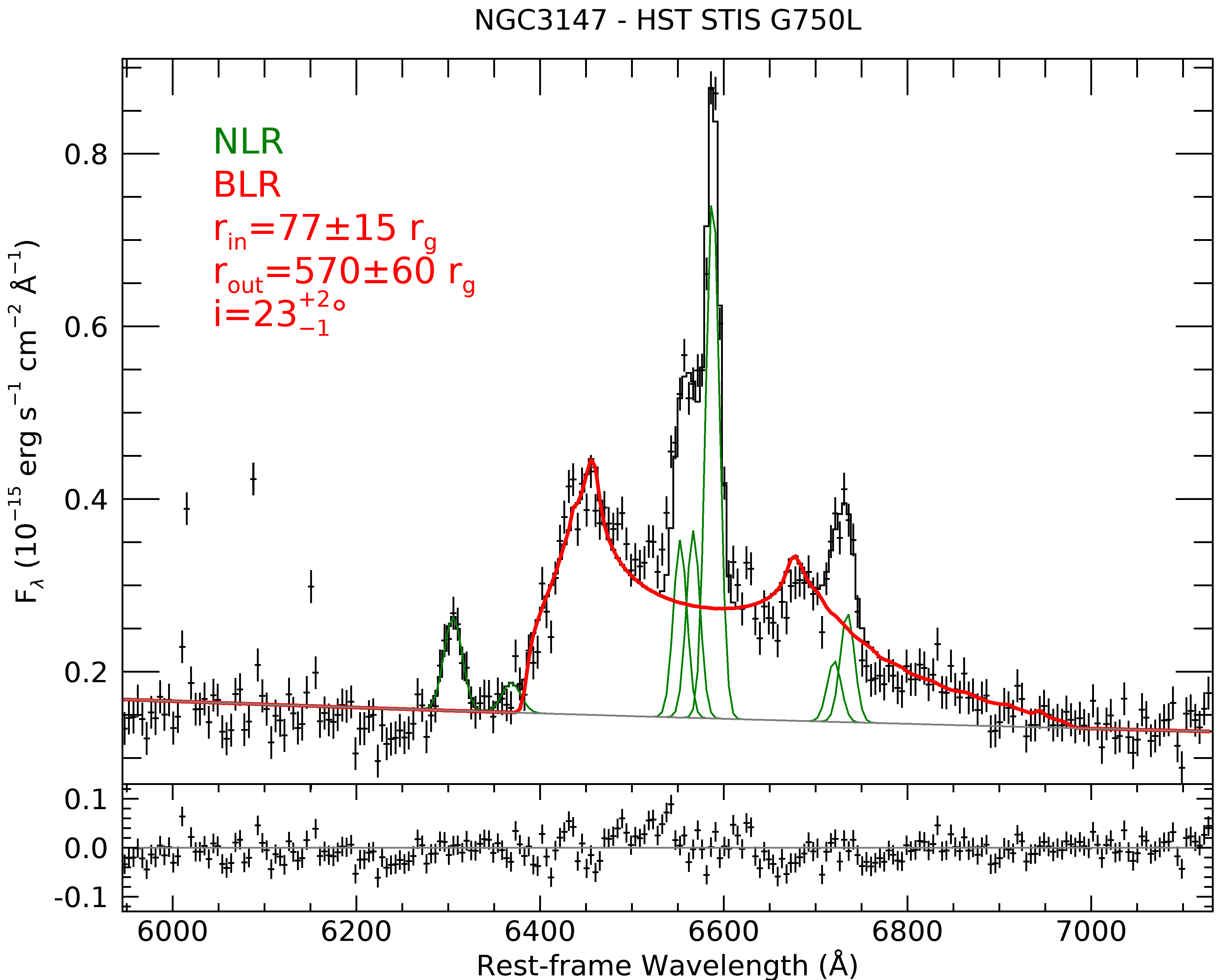}
\caption{\label{ngc3147_rel}A remarkably prominent and very broad H$\alpha$ line profile (in red) is revealed in the \textit{HST} STIS G750L spectrum of NGC~3147, carrying the signature of a relativistic thin Keplerian disk, with an inner emitting radius of only 77 $r_g$. Narrow lines from the NLR are in green, the continuum in grey. The fit residuals are shown in the bottom panel.}
\end{figure}

The \hst\ STIS G750L spectrum of NGC~3147 is shown in Fig.~\ref{ngc3147_rel}. Along with the lines from the narrow line region (NLR), it is evident the presence of a very broad emission line centred at the H$\alpha$ wavelength, extending from $\sim6450$ to $\sim7050$ \AA. This corresponds to a Full Width at Zero Intensity (FWZI) of $\sim27\,000$ km s$^{-1}$. 
We therefore focus our analysis in the $6000-7200$ \AA\ range, and started by modeling the narrow emission lines from the NLR. In our fits, we assumed that the wavelengths of the components of the [\ion{O}{i}], [\ion{N}{ii}] and [\ion{S}{ii}] doublets have the expected ratios $\lambda_2/\lambda_1 = 6363.81/6300.32$, $6583.39/6548.06$ and $\lambda_2/\lambda_1 = 6730.78/6716.42$, respectively, and share the same width. We also assumed that $\mathrm{F}([\ion{N}{ii}] \lambda6583.39)/\mathrm{F}([\ion{N}{ii}] \lambda6548.06) = 3$ and $\mathrm{F}([\ion{O}{i}] \lambda6300.32)/\mathrm{F}([\ion{O}{i}] \lambda6363.81) = 3$, as required by the ratio of the respective Einstein coefficients, and the width of the H$\alpha$ is the same as that of the [\ion{N}{ii}] doublet. The properties of the narrow emission lines modelled in the 6000-7200 \AA\ range are listed in Table~\ref{ngc3147_lines}.

\begin{table}
\caption{\label{ngc3147_lines}Emission line properties in the $6000-7200$ \AA\ range. Laboratory wavelengths (\AA) in air are from \citet{Bowen1960}. FWHMs are in km s$^{-1}$, fluxes in $10^{-15}$ erg cm$^{-2}$ s$^{-1}$.}
\begin{center}
\begin{tabular}{llll}
\multicolumn{4}{c}{\textsc{Narrow Emission Lines}}\\
\hline
Line & $\lambda_\mathrm{l}$ & FWHM & Flux \\
\hline
$\mathrm{[{O\,\textsc{i}}]}$ & 6300.32 & $1030^{+200}_{-160}$ & $2.7\pm0.4$\\[1ex]
$\mathrm{[{O\,\textsc{i}}]}$ & 6363.81 & $1030^{+200}_{-160}$ & $0.9^{+0.3}_{-0.6}$\\[1ex]
$\mathrm{[{N\,\textsc{ii}}]}$ & 6548.06 & $680\pm30$ & $3.90^{+0.13}_{-0.09}$\\[1ex]
H$\alpha$ & 6562.79 & $680\pm30$ & $4.1\pm0.3$ \\[1ex]
$\mathrm{[{N\,\textsc{ii}}]}$ & 6583.39 & $680\pm30$ &$11.7^{+0.4}_{-0.3}$\\[1ex]
$\mathrm{[{S\,\textsc{ii}}]}$ & 6716.42 & $520^{+150}_{-110}$ & $1.5\pm0.3$\\[1ex]
$\mathrm{[{S\,\textsc{ii}}]}$ & 6730.78 & $520^{+150}_{-110}$ & $2.2\pm0.3$\\
 &  &  &  \\
 \end{tabular}
\begin{tabular}{lllll}
\multicolumn{5}{c}{\textsc{Broad H$\alpha$ disk line profile}}\\
\hline
$i$ & r$_{\rm in}$ & r$_{\rm out}$ & $\alpha$ & Flux \\
\hline
$23^{+2}_{-1}\degr$ & $77\pm15$ r$_g$ & $570\pm60$ r$_g$ & $1.9\pm0.3$ & $67\pm2$\\
\end{tabular}
\end{center}
\end{table}

The residuals left after the inclusion of the narrow emission lines from the NLR show a very broad and asymmetric profile. In particular, it is characterized by a steep cutoff blue wing and an extended red wing, which are the signature of a mildly relativistic thin accretion disk line profile \citep[e.g.][]{Laor1991}.
We therefore adopted the \textsc{Kerrdisk} model \citep{Brenneman2006}, which reproduces the line profile from accretion disk systems around BHs. This model accurately reproduces the overall profile (Fig.~\ref{ngc3147_rel}), tightly constraining the emission parameters: the inclination angle is $i=23^{+2}_{-1}\degr$, the inner and outer radii are r$_{\rm in}=77\pm15$ and r$_{\rm out}=570\pm60$ r$_g$, respectively, where r$_g=GM/c^2$ is the gravitational radius.  The line emissivity index \citep[see][]{Brenneman2006} is constrained at $\alpha=1.9\pm0.3$, while the BH spin, as expected since r$_{in}$ is much larger than the innermost stable orbit, is completely unconstrained. The rest-frame line wavelength of the disk line component appears to be blueshifted with respect to the narrow H$\alpha$ line\footnote{Keeping the two wavelengths linked yields a statistically worse fit, but the other disk line parameters are only slightly affected, showing that the overall modelling of the profile is robust.} by $850^{+100}_{-200}$ km s$^{-1}$. It is interesting to note that all the derived parameters are linked to clear properties of the observed profile. The inclination angle is mostly set by the cutoff at lower wavelengths, while the inner radius is determined by the extension of the red wing tail. The distance between the two peaks, instead, is constrained by the outer radius.
The flux of the broad component of the H$\alpha$ emission line is $6.7\pm0.2\times10^{-14}$ erg cm$^{-2}$ s$^{-1}$, i.e. $\sim16$ times larger than the narrow H$\alpha$ line. This corresponds to a line luminosity of $1.30\pm0.04\times10^{40}$ erg s$^{-1}$, and an equivalent width of $449\pm13$ \AA, close to the mean value of 570 \AA\ observed in low z type 1 AGN \citep[e.g.][]{Stern2012}.

\begin{figure}
\includegraphics[width=\columnwidth]{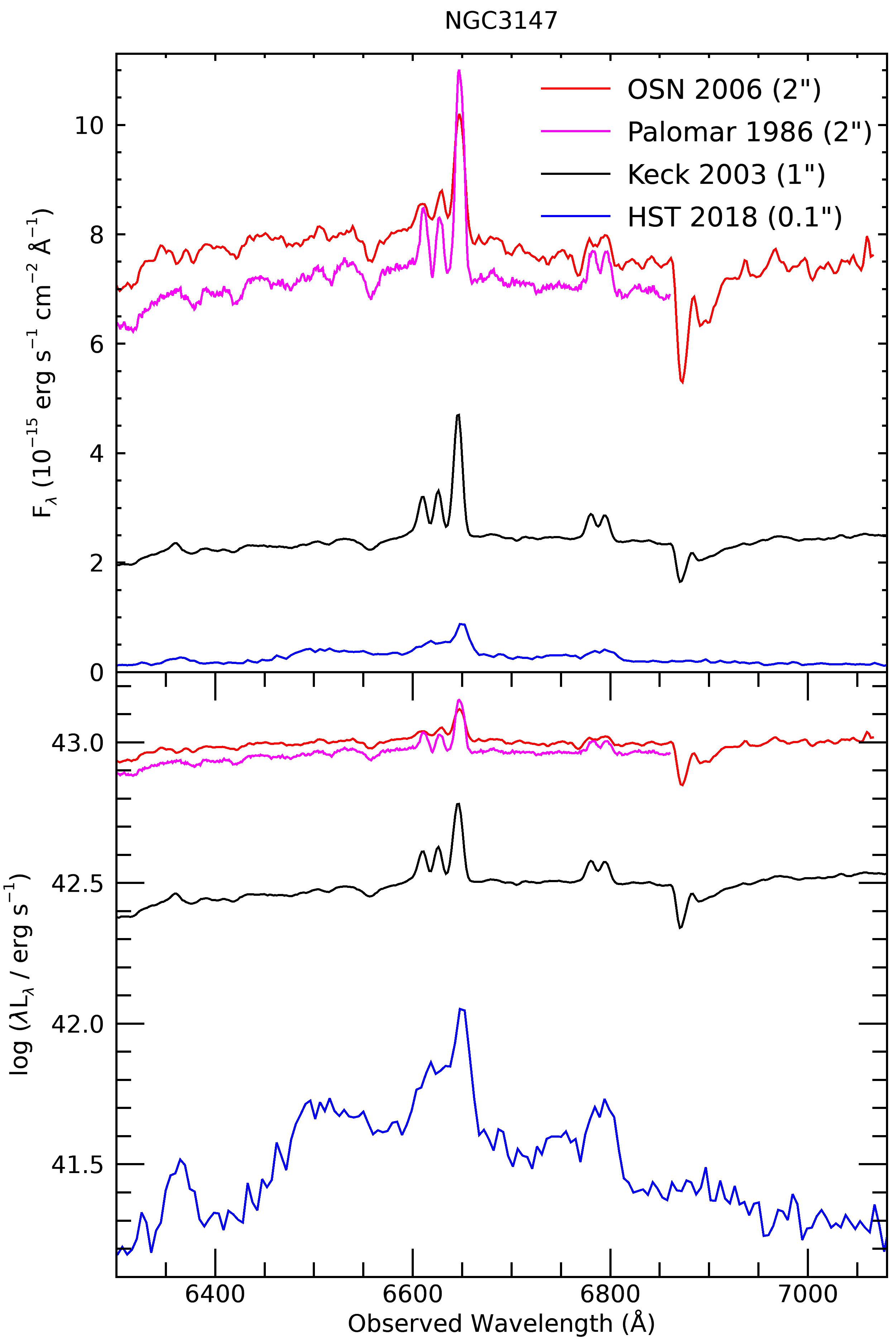}
\caption{\label{ngc3147_opt}Comparison of the available ground-based optical spectra of NGC~3147 with the \textit{HST} STIS G750L spectrum analysed here. The starlight in the $0.1\arcsec$ \textit{HST} spectrum is almost completely eliminated, revealing the AGN continuum, which is a factor of $\sim12$ smaller than the host light in the $1\arcsec$ Keck spectrum, and $\sim40-50$ smaller than the $2\arcsec$ Palomar and OSN spectra.}
\end{figure}

An important question is whether the broad component of the H$\alpha$ line was hidden by the host contamination in previous, ground-based spectra or is a new feature just arisen at the time of the \textit{HST} observation. Fig.~\ref{ngc3147_opt} shows on the same linear scale with the new \textit{HST} spectrum three ground-based spectra analysed in literature: the 1986 Palomar spectrum \citep{Ho1995}, the 2003 Keck spectrum \citep{Tran2011} and the 2006 Observatorio de Sierra Nevada spectrum \citep{Bianchi2008c}. The very small aperture of our \textit{HST} observation allowed us to exclude the starlight continuum and part of the NLR, which dominate the other spectra. Even the less contaminated (i.e. the highest S/N and smallest aperture) \textit{Keck} spectrum is compatible with the presence of the broad H$\alpha$ line observed by \textit{HST}. We have therefore no evidence in favour of a variability of this component, which is likely hidden by the host galaxy contamination in earlier observations. This scenario is consistent with the X-ray flux state of NGC~3147 in a TOO \textit{Swift} observation we triggered shortly after the \textit{HST} observation (July 27th): the 2-10 keV flux is $(1.5\pm0.5)\times10^{-12}$ erg cm$^{-2}$ s$^{-1}$, i.e. the average flux of the source \citep{Bianchi2017a}, thus excluding this is a `changing-look' AGN.

\section{Discussion}

Based on the analysis presented in this paper, it is unavoidable to conclude that NGC~3147 is \textit{not} a true type 2 AGN, but a low accretion rate type 1 AGN with a large BH mass. The BLR is clearly present in this source, and the previous lack of its detection was caused by the combined effect of a very broadened emission profile and an intrinsic weakness with respect to the host galaxy, as expected for a low accreting, high mass, object. Since NGC~3147 likely represented the best candidate for this class of sources, this result questions the very existence of true type 2 AGN.

Most models for the origin of the BLR predict its disappearance at low accretion rates, although for different reasons: the disk wind cannot form if the accretion rate is lower than the minimum value for which a standard \citet{Shakura1973} disk is stable \citep{Nicastro2000}; the photoionized gas in the inflated disk remains dusty, making the BLR emission inefficient \citep{Baskin2018a}. 
Inserting the stellar velocity dispersion $\sigma_*=233\pm8$ km s$^{-1}$ measured by \citet{VanDenBosch2015} in the $M-\sigma_*$ relation for ellipticals \citep{Gultekin2009}, we get a BH mass $\log M_{\rm BH}/M_\odot =8.49\pm0.11$ for NGC~3147 (the reported dispersion of the relation is 0.31 dex). Adopting the most recent X-ray bolometric corrections (e.g. Duras et al., in prep), a bolometric luminosity of $4.0\pm1.2\times10^{42}$ erg s$^{-1}$ can be derived from the average 2-10 keV X-ray luminosity \citep[$3.3\pm1.0\times10^{41}$ erg s$^{-1}$:][]{Bianchi2008c,Matt2012,Bianchi2017a}. Therefore, the Eddington ratio of NGC~3147 is $L_{\rm bol}/L_{\rm Edd}=1.0\pm0.4\times10^{-4}$. This accretion rate is much lower than the minimum value required by \citet{Nicastro2000} to launch the BLR wind ($1-4\times10^{-3}$, slightly depending on the BH mass and the accretion properties), and of the same order of the critical accretion rate below which dust effectively suppresses BLR emission, but only if low metallicity is allowed \citep{Baskin2018a}.

The luminosity of the broad H$\alpha$ line in the \hst\ spectrum ($\log L_\mathrm{b\mathrm{H}\alpha}=40.11$) is consistent with the predicted values of $40.5\pm0.5$ and $40.6\pm0.5$ from the $L_\mathrm{X}$ vs. $L_{\rm b\mathrm{H}\alpha}$ and $L_{[\ion{O}{iii}]}$ vs. $L_{\rm b\mathrm{H}\alpha}$ relations in type 1 AGN \citep{Stern2012,Stern2012c}, given the above reported X-ray luminosity and the (extinction corrected) [\ion{O}{iii}] luminosity of $1.6\times10^{40}$ erg s$^{-1}$\citep{Bianchi2008c}. NGC~3147 appears like a standard type 1 AGN at the very low end of luminosity ranges.

From the measured broad H$\alpha$ luminosity, we can also predict the BLR radius (as mapped by H$\alpha$) in this source. By combining the $L_\mathrm{b\mathrm{H}\alpha}$ vs. $L_{5100}$ and the $R_\mathrm{BLR}$ vs. $L_{5100}$ relations reported in \citet{Greene2005}, we get $R_\mathrm{BLR}=2.8\pm0.2\times10^{15}$ cm (the uncertainty being dominated by the uncertainties in the parameters of the relations). For the BH mass reported above, this translates into $R_\mathrm{BLR}=62^{+18}_{-14}$ r$_g$ (the uncertainty being dominated by the uncertainty on the BH mass). This is in very good agreement with the r$_{\rm in}=77\pm15$ r$_g$ found with the disk line profile fit.

The disk line profile fit yields the emitting radii in units of $r_g$, independently of the luminosity and the BH mass. We can therefore do the opposite reasoning, and estimate the BH mass in NGC~3147 on the basis of the inner radius and the luminosity of the H$\alpha$ line. By re-arranging and re-normalizing the above-mentioned formulae, we get

\begin{equation}
\log \frac{M_{\rm BH}}{M_\odot} =\left( 10.22\pm0.05\right) + \left(0.553\pm0.017\right) \log L_\mathrm{b\mathrm{H}\alpha}^{40} - \log \frac{\mathrm{r}_{\rm in}}{\mathrm{r}_g}
\end{equation} 

\noindent where $L_\mathrm{b\mathrm{H}\alpha}^{40}$ is the H$\alpha$ luminosity in units of $10^{40}$ erg s$^{-1}$. Inserting the values measured in our \hst\ spectrum, we have $\log M_{BH}/M_\odot =8.40\pm0.05$, in agreement with the $8.49\pm0.11$ obtained via the stellar velocity dispersion.

H$\alpha$ disk line profiles have been previously observed in a handful of objects \citep[e.g.][]{Eracleous1994,Eracleous2003,Strateva2003,Tran2010,Storchi-Bergmann2017}, but with significantly larger inner radii at least in the range $10^2-10^3\, \mathrm{r}_g$, the smallest being 98 r$_g$ in SDSS J0942+0900 \citep{Wang2005}.  The larger inner radii are indeed expected, since these objects have generally larger accretion rates than NGC~3147 and the results are commonly interpreted as a torus like structure, possibly associated with a geometrically thick low accretion rate configuration. Here we find that the BLR forms a thin disk, extending down well below 100 $r_g$. Standard BLR photoionization models require column densities $N_\mathrm{H}>10^{23}$ cm$^{-2}$. The disk gas mass within $R_\mathrm{BLR}$ is then $\pi R_\mathrm{BLR}^2 m_p N_\mathrm{H}> 2\times10^{-3}M_\odot$, which is $>3$ times the accretion required per year to sustain the observed $L_{bol}$. Therefore, this compact thin disk is likely the inner part of the reservoir of gas which feeds the AGN, i.e. the accretion disk. The presence of a thin accretion disk at $L/L_{\rm Edd}\sim10^{-4}$ is likely to change our current view of the accretion flow, 
being in contradiction with the standard paradigm, that at low accretion rates, the accretion configuration becomes optically thin and quasi-spherical \citep[][]{Blandford1999}. 

\begin{figure}
\includegraphics[width=\columnwidth]{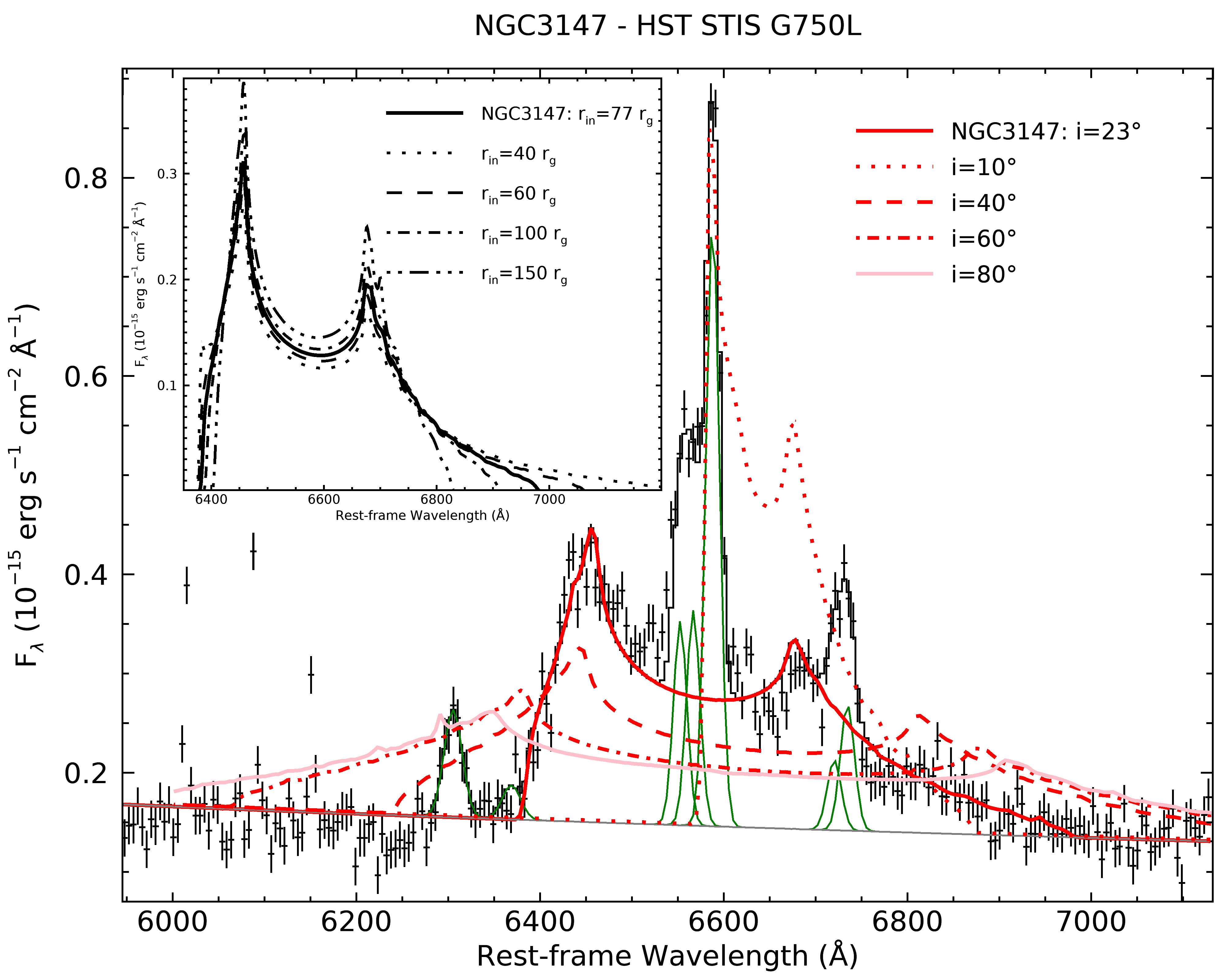}
\caption{\label{kerrdisk_i}Predicted disk line profiles with the same total flux and disk parameters as NGC~3147, but with different inclination angles. Even broader profiles are expected to be observed in higher inclination low luminosity AGN. The expected profiles for different inner radii are shown in the inset.}
\end{figure}

Even broader H$\alpha$ profiles are expected to be observed when an object like NGC~3147 is seen at a higher inclination (see Fig.~\ref{kerrdisk_i}). One should therefore look carefully for weak quasi continuum features
which can extend over a region $>1000$ \AA\ wide. Despite their weakness, these features should be clearly detected once the host contamination is fully excluded, and the net AGN emission is observed at a reasonable S/N ratio. Optical detections of relativistic line profiles may provide a new tool to explore the innermost disk structure: by exploring $L/L_{\rm Edd}\sim10^{-5}$ and $M_{\rm BH}\sim3\times10^9$ $M_\odot$ (typical in LINERs and FR Is) we may be able to go down by an additional factor of 10 (in terms of r$_g$) in the inner radius, and see the emission from the innermost disk at unprecedented resolution and S/N, as the optical photons detection rate is $10^5$ higher than in the X-rays. 

\section*{Acknowledgements}
We thank the nonymous referee and H. Tran for the 2003 Keck spectrum of NGC~3147. SB acknowledges financial support from the Italian Space Agency under grant ASI-INAF 2017-14-H.O. FJC acknowledges financial support through grant AYA2015-64346-C2-1P (MINECO/FEDER). 
The data described here may be obtained from the MAST archive at \url{http://dx.doi.org/10.17909/t9-7vzv-2j06}.
\bibliographystyle{mnras}
\bibliography{NGC3147HST}

\label{lastpage}

\end{document}